\documentclass[a4paper,11pt]{article}
\usepackage{pos}

\usepackage{physics}
\usepackage{adjustbox}

\usepackage[range-phrase=\textendash, range-units=single]{siunitx}
\let\sun\odot
\DeclareSIUnit\solarmass{\ensuremath{M_\sun}}
\DeclareSIUnit\erg{erg}
\DeclareSIUnit\year{yr}
\DeclareSIUnit\parsec{pc}

\title{Searching for sub-populations within the gamma-ray solar flares catalog: a graph-based clustering analysis}
 \ShortTitle{Searching for sub-populations within the gamma-ray solar flares catalog}

\author*[a]{Jonathan Mauro}
\author[a]{Gwenhaël de Wasseige}
\affiliation[a]{Centre for Cosmology, Particle Physics and Phenomenology - CP3, \\ Universite Catholique de Louvain, B-1348 Louvain-la-Neuve, Belgium}

\emailAdd{jonathan.mauro@uclouvain.be}
\emailAdd{gwenhael.dewasseige@uclouvain.be}

\abstract{Solar flares are highly energetic events that happen in the solar atmosphere. They are mostly observed as X-ray or gamma-ray bursts located on the Sun’s surface. While they are known to be sites of particle acceleration, the acceleration process(es) responsible for the observed fluxes remain unsure. The diversity in shape and duration of the $\gamma$-ray fluxes suggests the existence of distinct phases of hadronic acceleration. Moreover, different acceleration processes could explain the differences observed among flares. In this work we search for the evidence of sub-populations within the catalogue of gamma-ray solar flares observed by Fermi-LAT.
We aim at grouping flares with similar physical properties to be able to probe theoretical models for neutrino production within different classes of flares. We use measurements of the X-ray and $\gamma$-ray fluxes, as well as CMEs and SEPs, to cluster the events using a graph-based algorithm. Furthermore, we investigate the most representative features that characterise the identified sub-populations to allow for qualitative analysis and model development.
}

\ConferenceLogo{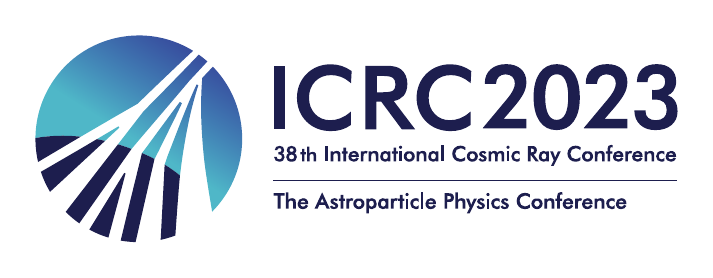}

\FullConference{
38th International Cosmic Ray Conference (ICRC2023)\\
  26 July - 3 August, 2023\\
  Nagoya, Japan}

\begin{document}
\maketitle

\section{Introduction}
\label{sec:introduction}

Solar flares are extremely energetic events that occur in the solar atmosphere, they are observed as peaks in the $\gamma$-ray flux and/or in the X-ray flux, and they are often associated with other interesting phenomena, such as Coronal Mass Ejections (CMEs) and production of Solar Energetic Particles (SEPs), i.e., high-energy charged particles. These observations prove solar flares can be the site of particle acceleration up to several GeV, and hence perfect candidates to be neutrino sources. Moreover, the link between gamma rays and neutrinos plays an important role, as discussed in \cite{gwen_thesis}: $\gamma$-ray observations and spectral analysis prove pion production to happen in the most energetic solar flares, implying also neutrino production in the MeV-GeV range, as a result of pion decay.
Solar flares have been mostly studied through the emissions of X-Rays and $\gamma$-rays. Mainly because of the measurements of the X-ray flux, the electron acceleration processes are quite well understood.

The high-energy $\gamma$-ray observations done by EGRET in the nineties, and later by Fermi-LAT, provided evidence for hadronic acceleration and highlighted different phases during flares \cite{1993A&AS...97..349K, Ajello:2021agj}. However, the observed shapes of  $\gamma$-ray fluxes are often not compatible with simple magnetic reconnection models used to explain the X-ray emissions and raise questions on acceleration processes. This directly reflects on the expected neutrino flux from solar flares, as neutrinos are mostly produced by pion decay, and, like $\gamma$-rays, they are a consequence of hadronic acceleration.

The history of the search for solar flare neutrinos begins in the late eighties when the Homestake experiment reported an increase in total neutrino flux in correspondence with a large solar flare \cite{Davis:1994jw}. While the validity of this result is still being discussed, many experiments, such as IceCube, KamLAND and Super-Kamiokande \cite{IceCube:2021jwt, KamLAND:2021sda, Super-Kamiokande:2022yrk}, have followed this path. Experimental evidence for solar flare neutrinos is yet to be found, however, their existence is theoretically guaranteed, and the difficulties that come with their detection make it a stimulating challenge for the neutrino astronomy community.

Besides the Homestake results, other experiments have joined the search for neutrinos from solar flares. Notably, Kamiokande II and LSD performed similar searches in the early nineties but didn’t report the same excess \cite{Hirata:1989mw, Aglietta:1991fv}. More recently, stringent constraints have been put on the solar flares neutrino flux by the KamLAND, SNO, and Borexino collaborations in the low MeV energy range \cite{KamLAND:2021sda, Aharmim:2013oba, Borexino:2019wln}. These measurements however still rely on the assumption that the neutrino flux is proportional to the X-ray flux as they look for coincidences in the time window around the X-ray peak. On the other hand, IceCube has performed a search in the GeV range by looking at $\gamma$-ray flares with significant pion decay signal observed by Fermi-LAT \cite{IceCube:2021jwt}, as these are expected to be the optimal population of neutrino emitters as discussed in \cite{gwen_thesis}.
Super-Kamiokande also followed with a similar analysis on the topic and put upper limits in the 20-110 MeV range, this search was performed using a mixed selection of $\gamma$-ray and X-ray flares as per availability of Fermi-LAT data \cite{Super-Kamiokande:2022yrk}. 

The relation between $\gamma$-ray and X-ray fluxes from solar flares is still being investigated and it is key to understanding the acceleration processes responsible for the observed emissions. Nonetheless, as the neutrino flux is expected to be proportional to the $\gamma$-ray one, this relation can be useful for flare selection in neutrino searches when gamma-ray measurements are not available. As discussed in \cite{Shih_2009}, there appears to be a close proportionality between the two fluences, however, this does not imply that same proportionality holds for short time windows, as highlighted by the more recent measurements performed by Fermi-LAT \cite{Ajello:2021agj}.

Both leptonic and hadronic acceleration processes are believed to be caused by the shock wave produced by magnetic reconnection following the reconfiguration of field lines in the solar atmosphere. Part of the accelerated particles are injected into the photosphere where they interact and produce X-rays, $\gamma$-rays and presumably neutrinos. However, shock acceleration can only explain the prompt emission phase, when the peaks in X-rays and  $\gamma$-rays are coincident, the delayed prolonged emissions observed in $\gamma$-rays imply a different hadronic acceleration process to take place, which doesn’t seem to produce relativistic electrons.
Because of the multi-messenger nature of solar flares, they represent the perfect source to study the acceleration process and to investigate the relations between electromagnetic emissions in different energy regimes, as well as the relative neutrino emissions. In particular, solar flares offer a unique opportunity to understand the hadronic processes leading to neutrino production as we have direct measurements of the $\gamma$-ray fluxes. In fact, $\gamma$-ray solar flares are not subject to the possible energy dispersion caused by the presence of an opaque medium between the source and the detectors, which instead characterises other transient $\gamma$-ray sources.

We want to stress here the importance of understanding the intrinsic structure of the solar-flare population. Extrapolating information from a diverse sample of flares might be misleading, and we believe that the possibly different underlying acceleration processes highlighted in the Fermi-LAT catalog \cite{Ajello:2021agj}, might prevent us to draw correct conclusions on the behaviour of a `typical flare'.
In particular, this is relevant for the way in which the flares, and respective time windows, are selected for neutrino searches.

\section{Method}
\label{sec:method}

In order to identify the inherent substructure of our data we make use of unsupervised machine learning techniques. Depending on the dataset's properties (e.g., dense, sparse, normally distributed), different clustering algorithms are best suited to identify sub-population.
Hierarchical Clustering Algorithms (HCAs) are mostly used when dealing with different-size clusters and are preferred when dealing with small datasets. As opposed to other algorithms that find clusters by sectioning the parameter space, HCA investigates the distances between data points, so it can be implemented using any distance (or similarity) metric of choice.

\subsection{Solar-flare dataset}
We define a flare in our analysis by its observables that are connected to particle acceleration. In addition to the fluence and duration recorded by Fermi-LAT, we have used observations in the X-ray from GOES (and STEREO for behind-the-limb flares) satellites as well as CME measurements. For this analysis, we select 34 solar flares that are detected by the Fermi-LAT SunMonitor and that have associated CMEs and SEPs measurements (These flares can be found in Table 1 and Table 4 of the Fermi-LAT paper \cite{Ajello:2021agj}). The various observables we have used are the GOES Class, the duration of the GOES flare and the Fermi-LAT flare, the CME speed, the maximum energy (emax) detected in SEPs and Hard X-Ray (HXR). For detailed information about these quantities, we refer the interested reader to the original paper. In the following, each solar flare will be represented as a vector of these observables.
\subsection{Clustering algorithm}

\begin{figure}[h]
    \centering
    \includegraphics[width=0.8\textwidth]{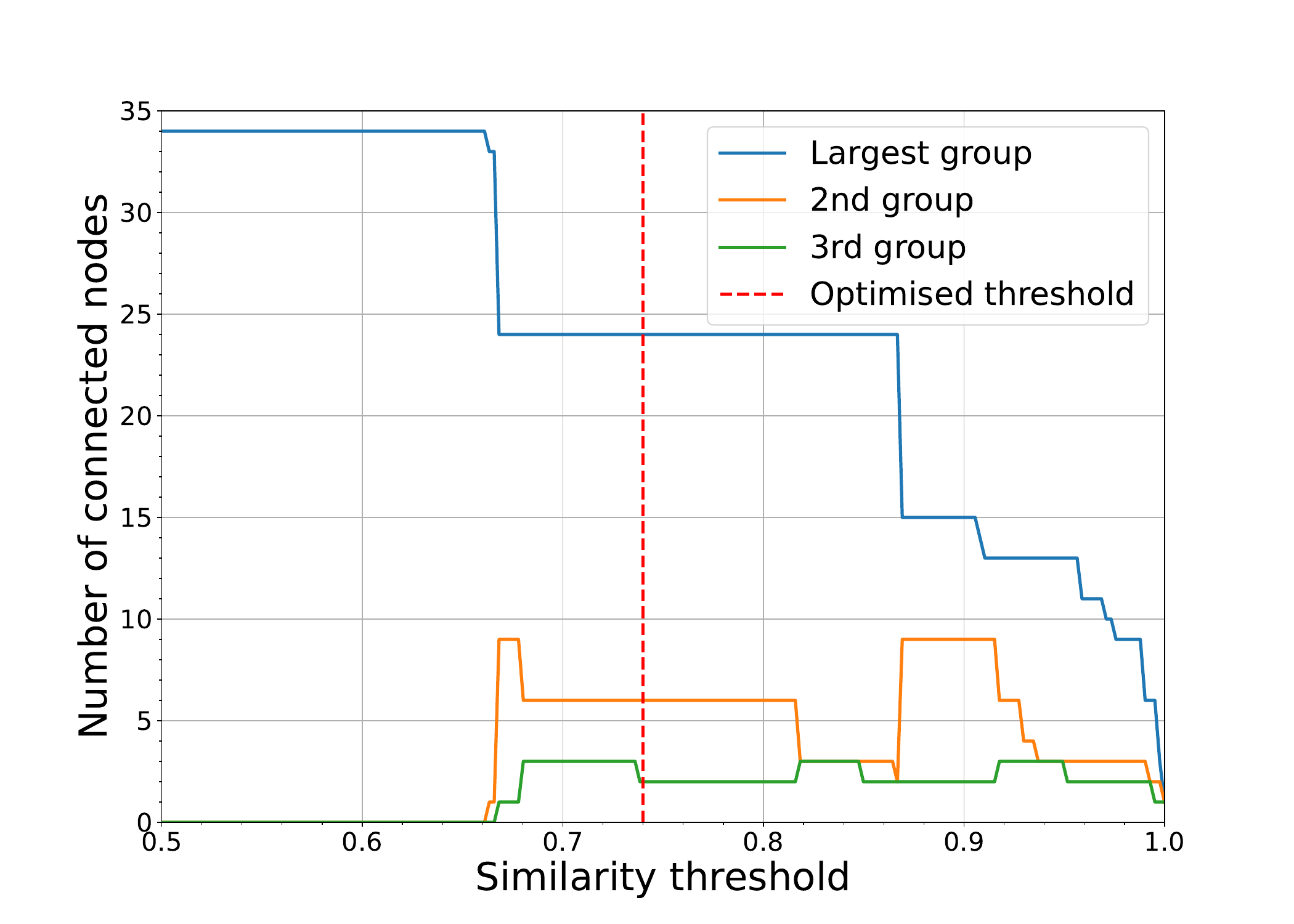}
    \caption{Number of nodes in the three largest clusters as a function of the threshold on the cosine similarity between nodes. \label{fig:thresholds}}
\end{figure}

The clustering analysis that we perform is based on a graph representation of clusters and is a variation of a more traditional single linkage HCA. The main difference is that our method uses similarities between individual data points, while traditional HCAs measure the distance between clusters in order to merge them. In the graph representation, this results in the clusters being defined by highly connected graphs. Compared to other HCAs, this method allows for a better understanding of the cluster substructure.

In the method we propose, each solar-flare event is represented as a node. We define a similarity function in the parameter space, which describes the pairwise similarity between flares. An edge is drawn between two nodes if the similarity between the events is above a certain threshold. In such a way, for a high enough threshold, we will have a number of clusters k equal to the number of nodes or flares, while for a similarity threshold close to zero we would have $k = 1$ corresponding to one fully connected graph to which all of the flares belong.

The steps to our clustering algorithm are the following:
\begin{itemize}
    \item As per common practice, we start by scaling our dataset, i.e. we reshape each variable's distribution to have mean $\mu = 0$ and unit variance;
    \item We use Principal Component Analysis (PCA) to reduce the dimensionality of our data;
    \item We build a similarity matrix by computing the pairwise cosine similarity between our data in the Principal Components (PCs) space;
    \item We map our data to nodes into a graph and draw edges between flares that have a similarity above a threshold;
    \item We define the cluster to be the connected subgraphs;
    \item Finally, we optimise for a value of the threshold that produces the most stable configuration in the clusters.
\end{itemize}

We reduce the original 7-variables dataset to three PCs. As no clear hierarchy could be seen in the explained variance of the PCs, the number of components was picked to conserve at least 75\% of the total variance.
The correlation of the PCs to the original variables is shown in Table~\ref{tab:parameters}.

\begin{table}[]
\centering
\caption{Correlation coefficient between the original observables considered in our analysis (first line) and the new parameters that are obtained using PCA. The highest coefficients are highlighted in bold. \label{tab:parameters}}
\begin{adjustbox}{width= 0.72\paperwidth}
\begin{tabular}{|l|l|l|l|l|l|l|l|}
    \hline
            & GOES class     & GOES duration     & Fermi-LAT duration  & CME speed     & GR fluence    & SEP emax      & HXR emax     \\ \hline
    PC1     & ~0.68     & ~0.21         & ~0.82     & \textbf{~0.84}& \textbf{~0.83}& ~0.57         & ~0.53        \\ \hline
    PC2     & ~0.47     & \textbf{-0.88}& -0.20     & -0.28         & ~0.08         & -0.14         &\textbf{~0.54}\\ \hline
    PC3     & ~0.24     & -0.22         & -0.16     & ~0.05         & -0.29         & \textbf{~0.73}&\textbf{-0.37}\\ \hline
\end{tabular}
\end{adjustbox}
\end{table}

Our pairwise similarity function of choice is the so-called `cosine similarity', which is a common choice in clustering algorithms. It returns values from -1 to 1 and is defined as

\begin{equation}
    S_{\cos}(\vb{X}, \vb{Y}) = \frac{\langle{\vb{X}, \vb{Y}}\rangle}                  {\norm{\vb{X}}\norm{\vb{Y}}},
\end{equation}
where $\vb{X}$ and $\vb{Y}$ are two generic vectors, and in our case these would be flares in the PCs space.

To find the optimal threshold, we search for a stable configuration of the clusters' size. As shown in Fig.~\ref{fig:thresholds}, we investigate how the size of the three biggest clusters in our data changes as a function of the threshold. The threshold is found to be 0.74, and corresponds to the smallest value that intersects the longest plateaux for the three clusters, i.e., we maximise the sum of the lengths of the intersected plateaux over the three biggest clusters. With this method we expect the distances between nodes of separate clusters to be larger than the typical distances within the nodes of the same cluster.

\section{Results}
\label{sec:results}

\begin{figure}[]
    \centering
    \includegraphics[width=0.7\textwidth]{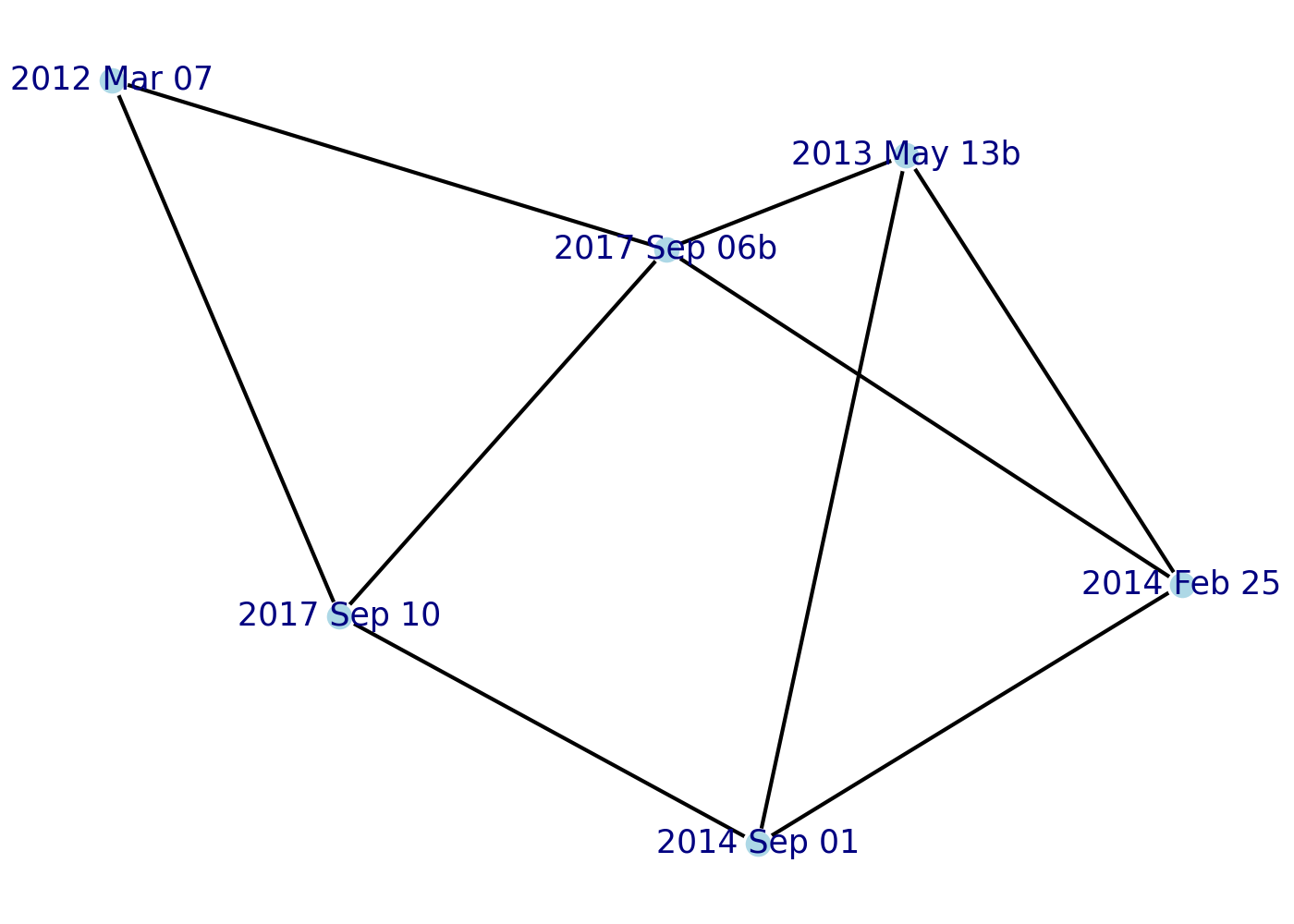}
    \caption{Graph representation of the second biggest cluster, the labels show the flares corresponding to each node.\label{fig:graph}}
\end{figure}

Within the 34 selected Fermi-LAT flares we find three clusters and two outliers, i.e., flares that do not belong to any of the clusters.
These two outliers are the solar flares occurring on 2011-08-09 and 2012-01-27. The 2011-08-09 Fermi-LAT flare shows a clear prompt emission and is most likely associated with a GOES X-Ray flare of class X6.9, and a CME with speed $1610~\text{Km}~\text{s}^{-1}$. On the other hand, the 2012-01-27 Fermi-LAT has a distinct delayed emission and no detected prompt component. It is most likely associated with a GOES X-Ray flare of class X1.7, and a CME with speed $2508~\text{Km}~\text{s}^{-1}$. It is worth mentioning that the 2012-01-27 flare is found close to the smallest cluster, and in fact, for a similarity threshold just below the optimised value, it connects to the corresponding subgraph. It can be seen from Fig.~\ref{fig:thresholds} that such a lower threshold would also produce a relatively stable configuration of cluster sizes. The opposite is true for the 2011-08-09 flare, in fact, such flare has significantly lower similarity with all of the others, and for increasing values of threshold, it is the first flare to be found disconnected from the biggest cluster.

The smallest cluster we find is composed of the 2012-05-17 and 2014-01-06 flares. The former is associated with an M5.1 GOES flare and a CME with speed $1582~\text{km}~\text{s}^{-1}$, while the second is an extremely energetic flare happening behind the limb of predicted X3.5 class and a CME with speed $1402~\text{km}~\text{s}^{-1}$.

The second biggest cluster contains 6 flares, and its graph representation is shown in Fig.~\ref{fig:graph}. Notably, all flares with $\gamma$-ray fluence above 10 cm$^{-2}$ belong to this cluster. Moreover, the 2017-09-06b flare, which is also found in this cluster, is associated with an X9.3 GOES flare, which is the highest class present in the Fermi-LAT catalog. The 2017-09-10 flare, the brightest solar flare detected by Fermi-LAT so far, is also present in this cluster.

The remaining 24 flares belong to the biggest cluster. The subgraph corresponding to this cluster is densely connected and we find in it a large clique (i.e., an induced complete subgraph) that comprises 11 flares. The flares belonging to this clique are the following: 2011-08-04, 2011-09-07, 2011-09-24, 2013-04-11, 2013-05-13a, 2013-05-15, 2013-10-25, 2013-10-28, 2014-06-10, 2014-06-11, 2017-09-06a. This clique consists of all of the X-ray flares with a duration smaller than 2500~s within the cluster, with the only exception being the short flare of 2011-09-06.

\section{Conclusions}
\label{sec:conclusion}

We discuss here the motivation for a sub-population search within solar-flare catalogs. We highlight the differences that are found between the X-ray flares observed by GOES, and the $\gamma$-ray Fermi-LAT flares. In particular, we stress how the presence of sub-populations of flares could affect solar-flare neutrino searches.

We select a sample of 34 flares from the Fermi-LAT catalog and we use the $\gamma$-ray fluence and durations together with measurements of associated X-ray flares, CMEs, and SEPs to build our dataset. We investigate the substructure of this dataset using a novel clustering algorithm which is based on single linkage HCA and graph representation of clusters.

The algorithm that we present makes use of PCA to reduce the dimensionality of the data, and a pairwise cosine similarity to create a similarity matrix. Flares are then mapped to nodes in a graph where edges connect nodes with similarity above an optimized threshold of 0.74.
Clusters are identified as the connected components of the graph.

Among the selected flares, we find 3 clusters and two outliers. We investigate the observables that unite the flares, and we find the most energetic $\gamma$-ray flares and the highest class GOES flare in the same cluster. We obtain a large cluster of 24 flares, and by investigating its graph structure we find a clique of 11 nodes corresponding to short-duration X-ray flares.

Future efforts will focus on implementing the solar-flare dataset with more events and observables, as well as optimising the clustering algorithm.

\bibliographystyle{JHEP}
\bibliography{references}

\end{document}